\documentclass[a4paper]{article}
\usepackage[vcentering,dvips]{geometry}
\usepackage{epsfig}

\usepackage{epic}
\usepackage{amsmath}
\usepackage{amsfonts}
\usepackage{amssymb}
\usepackage{amsthm}
\usepackage[mathscr]{eucal}
\usepackage{amsmath}

\usepackage[english]{babel}
\usepackage[iso-8859-7]{inputenc}
\usepackage[T1]{fontenc}
\usepackage{fancyhdr}
\fancypagestyle{plain}{%
\fancyhead{}
}
\RequirePackage{ifthen}
\RequirePackage{calc}
\RequirePackage{eso-pic}
\RequirePackage{graphicx}

\usepackage[perpage,hang]{footmisc}

\usepackage{gray}
\usepackage{quantum}
\usepackage{graphicx}
\usepackage{longtable}

\newlength\boxwidth

\begin{document}

\definecolor{light}{gray}{.75}
\def\GRAMMH{\noindent{%
\color{light}
\hrule width 126mm height 1mm
\color{light}
} }

\noindent
{\large\bfseries 
Quantum mechanical potentials
exactly solvable

\noindent
in terms of higher hypergeometric functions. I:

\noindent
The third-order case

}

\smallskip

{\narrower

\noindent
{\footnotesize 
S. Trachanas

\noindent
\hangindent=-3cm \hangafter=0
Foundation of Research and Technology Hellas (FORTH)\hfill\break
and Department of Physics, University of Crete, Crete, Greece

} }

\medskip

\noindent
\hangindent=-1cm \hangafter=0
{\narrower
{\footnotesize
{\bf Abstract}.\
We present a new
six-parameter family of potentials whose
solutions are expressed in terms of the hypergeometric
functions ${_3}F_2$, ${_2}F_2$ and ${_1}F_2$. Both the 
scattering data and the bound states of these potentials
are explicitly computed and the peculiar properties of the 
discrete spectrum are depicted in a suitable phase diagram. 
Our starting point is a third-order formal eigenvalue equation
of the hypergeometric type (with a suitable solution known)
which is transformed to the {Schr\"{o}dinger} equation
by applying the reduction of order technique as the crucial first step.
The general preconditions allowing for the reduction to 
{Schr\"{o}dinger} form of an arbitrary eigenvalue
equation of higher order, are discussed at the end of the 
article, and two universal features of the potentials arising
this way are also stated and discussed. In this general scheme the
{Natanzon} potentials are the simplest special case, 
those presented here the next ones, and so on for potentials 
arising from equations of fourth or higher order.

}}

\bigskip
\bigskip
\noindent
{\footnotesize\bfseries Ι. INTRODUCTION} 

\medskip


The search for exactly solvable quantum mechanical potentials has a long
history dating back to the early days of quantum mechanics. In 1971, 
Natanzon$^1$ appeared to have closed the subject by explicitly 
constructing the full set of potentials $V(x)$ whose solutions can be expressed
in terms of the hypergeometric functions $_2F_1$ and $_1F_1$.

The subject received a new impetus by the addition to the available tools 
for exact solution of the {\em {Darboux} transformation}$^2$:
A mapping of the {Schr\"{o}dinger} equation to itself which can 
be applied to any solvable potential and produce an infinite chain of 
new ones, usually with predetermined spectral changes with respect to the 
initial potential (addition or removal of bound states, etc.).

Given though that the solutions of these "derived potentials" are also 
expressed in terms of the hypergeometric functions 
$_2F_1$ and $_1F_1$ -- or, to be precise, in terms of linear combinations
of these functions and their derivatives$^3$-- it is reasonable to view these
potentials as simple "derivatives" of the initial ones. So, a complete 
description of the now-known exactly solvable potentials is, we believe, 
this: {\em   Exactly solvable are the {Natanzon} potentials 
and their {Darboux} derivatives}. 

In view of the above, it is clear that new solvable potentials can be 
discovered only if we enlarge the set of functions within which their 
solution is sought for. And the most natural such enlargement is 
the full set of hypergeometric functions $_pF_q$. These are certainly
equally exact functions as $_2F_1$ or $_1F_1$, the only difference being
that they satisfy linear differential equations of order higher than second. 
Specifically, the function $_pF_q$ satisfies an equation of order 
$q+1$ where for any given $q$ the index $p$ takes the values 
$0\leqq p\leqq q+1$ characterizing the various types of hypergeometric 
functions of the given order. 

In particular, if we restrict ourselves to hypergeometric functions of 
the third order --e.g., the functions $_3F_2$, $_2F_2$, $_1\!F_2$ and $_0F_2$--
then it is clear that their 
``use'' for constructing new exactly solvable potentials
presupposes,
as a necessary first step, the {\em reduction of order} of the third-order
equation satisfied by these functions. But it is also necessary that this 
initial third-order equation has an {\em eigenvalue parameter} $\lambda$
suitably located so that it takes the position of energy in the final 
{Schr\"{o}dinger} equation produced by the reduction. How this can be
done will be examined in the following section. 


\medskip
\bigskip
\noindent
{\footnotesize\bfseries {II}. THIRD-ORDER EIGENVALUE EQUATIONS
REDUCIBLE TO THE {SCHR\"ODINGER} FORM.}
\smallskip

As is well-known, if a particular solution $y=y_0$ of a linear differential 
equation is available, the transformation $y=y_0Y$ --that is, the factoring
out of the known solution-- eliminates the $Y$ term of the new equation and 
so the further substitution $Y'=u$ lowers its order by one. In the case
of a linear equation of order three

\begin{equation}\label{ENA}
a(x)y'''+b(x)y''+c(x)y'+d(x)y=0
\end{equation}
the application of this procedure leads to the second order equation

\smallskip
\begin{equation}\label{DYO}
u''+\left({b\over a}+3{y_0'\over y_0}\right) u'+
\left({c\over a}+2{b\over a}\,{y_0'\over y_0}+3{y_0''\over y_0}\right) u=0,
\end{equation}

\smallskip
\noindent
where $u=(y/y_0)'$. 
With the further substitution 

\smallskip
\begin{equation}\label{TRIA}
u(x)=y_0^{-3/2}\exp \left( -{1\over 2}\int {b\over a} dx\right) {\cal U} (x)
\end{equation}

\smallskip
\noindent
equation \eqref{DYO} is transformed into the {\em canonical form}

\smallskip
\begin{equation}\label{TESSERA}
{\cal U}''+\left({c\over a}-{1\over 2}\left({b\over a}\right)'
-{1\over 4}\left({b\over a}\right)^2+{1\over 2}\,{b\over a}\,{y_0'\over y_0}
+{3\over 2}\left({y_0'\over y_0}\right)'+{3\over 4}
\left({y_0'\over y_0}\right)^2\right) {\cal U}=0
\end{equation}

\smallskip
\noindent
where the first derivative term is missing. From \eqref{TESSERA} it is now 
clear that it can be cast into the so-called  {\em {Liouville} form}
\begin{equation}\label{PENTE}
{\cal U}''+\big(\lambda w(x)-v(x)\big){\cal U} =0 
\end{equation}
--which is readily reduced to the {Schr\"{o}dinger} equation--
if the following conditions are met: i) The coefficients $a(x)$, $b(x)$ 
and the special solution $y=y_0$ of \eqref{ENA}, are independent of the 
eigenvalue parameter $\lambda$, ii) the coefficient $c$ depends linearly on 
$\lambda$. That is, 

\smallskip
\begin{equation}\label{EXI}
c(x,\lambda) =c_0(x)+\lambda c_1(x) .
\end{equation}

\smallskip
\noindent
If these conditions are met then \eqref{TESSERA} is indeed a Liouville equation 
with {\em weight function} $w(x)=c_1(x)/a(x)$
and {\em {Liouville} potential} 
--this is a suitable name for the function $v(x)$ in \eqref{PENTE}-- as follows
\begin{equation}\label{EPTA}
v(x)=-{c_0\over a}+{1\over 2}\left({b\over a}\right)' +
{1\over 4}\left({b\over a}\right)^2-{1\over 2}\,{b\over a}\,{y_0'\over y_0}
-{3\over 2}\left({y_0'\over y_0}\right)'
-{3\over 4}\left({y_0'\over y_0}\right)^2 .
\end{equation}
As far as equation \eqref{PENTE} is concerned it is well known that it can be reduced to the {\em {Schr\"{o}dinger} form}
\begin{equation}\label{AALPHA}
\psi''(z)+\big(\lambda -V(z)\big)\psi (z)=0 
\end{equation}
--where primes stand now for derivatives with respect to $z$--
with the so-called {\em {Liouville} transformation}

\smallskip
\begin{equation}\label{OKTO}
\psi (z)=\psi\big( x(z)\big) =w^{1/4} {\cal U} (x)\big|_{x=x(z)}
\end{equation}

\smallskip
\noindent
where the {\em final position variable} $z$ is defined by 
\begin{equation}\label{ENNIA}
z'(x)=\sqrt{w(x)}\;\;\Rightarrow\;\; z(x)=\int\sqrt{w(x)}\,dx
\end{equation}
and the potential $V(x)$
--from now on we will always assume that $x=x(z)$-- is given by
\begin{equation}\label{DEKA}
V(x)={v(x)+\{ z, x\}\over w(x)}, 
\end{equation}
where $\{ z, x\}$ is the so-called {\em Schwartzian derivative} 
--or simply the {\em Schwartzian}-- of the function $z(x)$ defined by
\begin{equation}\label{ENTEKA}
\{ z, x\} = {1\over 2}\left({z''\over z'}\right)' -{1\over 4}
\left({z''\over z'}\right)^2={1\over 4}\left({w'\over w}\right)'
-{1\over 16}\left({w'\over w}\right)^2.
\end{equation}
It is now clear from \eqref{ENTEKA} that the integration constant in 
\eqref{ENNIA} is eliminated and can be ignored. Concerning the requirement
that the special solution $y_0$ is independent of $\lambda$ 
--in spite of the fact that $\lambda$ appears in the equation--
this leads also to the condition that the coefficient $d$ in \eqref{ENA} is 
linearly dependent on $\lambda$ like $c$. That is, 
\begin{equation}\label{DODEKA}
d(x,\lambda) =d_0(x)+\lambda d_1(x),
\end{equation}
whence equation \eqref{ENA} can finally be written as
\begin{equation}\label{DEKATRIA}
(L+\lambda M) y=0,
\end{equation}
where the operator $L$ has the full order of the equation
--that is, {\em three} -- 
while $M$ is of first order.
We claim therefore that any {\em formal}
--i.e., with no boundary conditions specified-- 
eigenvalue equation of the form 
\eqref{DEKATRIA} can be reduced to the {Schr\"{o}dinger} equation,
provided that the solution $y=y_0(x)$ of the first-order equation $M y=0$
is also a solution of $Ly=0$, and therefore also of the complete equation 
\eqref{DEKATRIA} for all values of the parameter $\lambda$.
A suitable generalization of this proposition for equations
of arbitrary order will be given in section VII.

\bigskip
\medskip
\noindent
{\footnotesize\bfseries ΙΙΙ. EXACTLY 
SOLVABLE EIGENVALUE EQUATIONS OF THIRD ORDER AND  
THE ASSOCIATED POTENTIALS}
\medskip

The next step is obvious. Starting with the most general third-order linear
equation, solvable in terms of hypergeometric functions, we write it in the
form  \eqref{DEKATRIA}
--with its parameters restricted so that the solution of $M y=0$
is also a solution of $Ly=0$-- and then transform it into 
{Schr\"{o}dinger} form producing along the way the respective 
solvable potential. The most general third-order equation we need is 
written as 
\begin{equation}\label{DEKATESSERA}
x^2(1+\omega x)y'''+x(\alpha +\beta x)y'' +
(\gamma +\delta x)y'+\left({\varepsilon\over x}+\zeta\right) y=0
\end{equation}
whose form is readily recognizable if we introduce the concept of 
{\em dimension} $d=m-n$ of the typical term 
$x^my^{(n)}$ (of an arbitrary linear equation with polynomial coefficients) 
and say that 
\eqref{DEKATESSERA} is a {\em bidimensional equation}.
That is, an equation whose terms can be grouped in only two different sets, 
each set having a definite dimension. Bidimensional equations are 
important because their solution in power series leads to a {\em two-term 
recursion formula} --e.g., $a_{k+\ell}=f(k)a_k$--
that allows us to determine the general series coefficient 
 $a_k$ in {\em closed form} as a function of $k$.

Thus the series represents an {\em exactly known function}
and the pertinent equation can be classified as
{\em exactly solvable} on purely mathematical grounds.
Furthermore, since the dimension $d$ of the typical term $x^ny^{(n)}$
is but the {\em displacement} on the exponent of a typical 
power $x^k$ brought about by the operator $L=x^m\partial^n$,
it follows that for a bidimensional equation with dimensions
$d_1$ and $d_2$, the integer $\ell$ in the recursion relation
$a_{k+\ell} =f(k)a_k$ 
\hbox{--which} is clearly equal to the difference of the two
displacements experienced by the general power of the series-solution
when substituted into the equation-- will be 
 $\ell =d_2-d_1$ ($d_2>d_1$ by convention). The quantity $\ell$ 
will be called the {\em step} of the equation since it indeed tells us
that the series-solution proceeds in steps of size $\ell$.
Based on the preceding discussion, Eq. \eqref{DEKATESSERA}
is uniquely specified by the statement that:
{\em It is the most general bidimensional equation of third order
and step unity}.

Concerning our assertion that Eq. \eqref{DEKATESSERA} is solvable in terms of
hypergeometric functions, this follows directly from the existence of a
two-term recursion relation and will be discussed further in 
Appendix A where we will also provide a simple recipe for arriving at the 
solution using only elementary algebraic operations. Note also that the 
exponents of power behaviors at zero and infinity, for any bidimensional 
equation, are determined by the conditions

\begin{equation}\label{llaa}
L_1x^{\mu}=L_1(\mu)x^{\mu +d_1}=0,\qquad L_2x^{\nu}=L_2(\nu)x^{\nu +d_2}=0
\end{equation}
i.e., from the roots of the {\em characteristic polynomials} 
$L_1(\mu)$ and $L_2(\nu)$ of the {\em unidimensional operators} $L_1$ and $L_2$
representing the components with the {\em least} and the {\em largest-dimension}
respectively, of the bidimensional operator $L=L_1+L_2$.

From the preceding discussion it follows that our starting point
should be the solvable, third-order eigenvalue equation 
$$x^2(1+\omega x)y'''+x(\alpha +\beta x)y''+(\gamma +\delta x)y'
+ \left({\varepsilon\over x}+\zeta\right)y +$$
\begin{equation}\label{DEKAENNIA}
\lambda\left( (\overline{\gamma}+\overline{\delta}x)y'+
\left({\overline{\varepsilon}\over x}+\overline{\zeta}\right) y\right) =0
\end{equation}
which has the same 
form as eq. \eqref{DEKATRIA} with
\begin{equation}\label{EIKOSI}
L=x^2(1+\omega x)\partial^3+x(\alpha +\beta x)\partial^2 +
(\gamma +\delta x) \partial
+ \left({\varepsilon\over x}+\zeta\right) ,
\end{equation}
and
\begin{equation}\label{EIKOSIENA}
M=(\overline{\gamma}+\overline{\delta}x) \partial +
\left({\overline{\varepsilon}\over x}+\overline{\zeta}\right)
\qquad (\partial =d/dx) .
\end{equation}
where $\overline{\gamma}$, $\overline{\delta}$, $\overline{\varepsilon}$
and $\overline{\zeta}$ are new, arbitrary parameters, independent of 
 $\gamma$, $\delta$, $\varepsilon$ and $\zeta$ of Eq. 
 \eqref{EIKOSI} but with the same role as these, 
hence the similar notation. 
With $M$ given by the expression above, the general solution of 
$My=0$ takes the form
$y=y_0=x^s(x+\rho)^{s'}$
--with suitably chosen $s$, $s'$ and $\rho$--
but here we will restrict our discussion to the case
$s'=1$ since then $y_0$ can certainly be also a solution of 
$Ly=0$ as it is a {\em terminating series} about $x=0$ 
and the equations of hypergeometric type are especially these
admitting such solutions. 

By regarding now $s$ and $\rho$ as {\em given parameters}
and requesting that $y_0=x^s(x+\rho)$ be a simultaneous solution
of the equations $My=0$ and $Ly=0$, we find that the operator $M$
can only have the form
\begin{equation}\label{EIKOSIDYO}
M=\left( 1+{x\over \rho}\right) \partial -\left({s\over x}+{s+1\over \rho}\right)
\end{equation}
while $L$ has the form of \eqref{EIKOSI}, with the parameters
$\delta$, $\varepsilon$ and $\zeta$ no longer independent, but 
expressed in terms of $\alpha$, $\beta$, $\omega$, $\rho$
and $s$ as follows:
\begin{equation}\label{EIKOSITRIA}
{\displaystyle{\delta  ={1\over\rho}
\big( 3(1-\omega\rho)s^2+(2\alpha
-2\beta\rho +3\omega\rho -3)s\big)}\hfill}\atop
{{\displaystyle{\varepsilon  
=-\big( s^3+(\alpha -3)s^2+(2-\alpha)s\big)}\hfill}\atop
{\displaystyle{\zeta  = - {1\over\rho}
\big( (3-2\omega\rho)s^3+(2\alpha -\beta\rho)s^2 +
(2\alpha -\beta\rho +2\omega\rho -3)s\big)} .}} 
\end{equation}
As for the parameter $\gamma$ it can be set equal to zero
since it only appears as an additive constant to the potential and 
can therefore be ignored. 

Our assertion becomes now quite specific: 
We claim that the formal eigenvalue equation of third order
\begin{equation}\label{EIKOSITESSERA}
x^2(1+\omega x)y'''+x(\alpha +\beta x)y''+
\left(\delta x+\lambda\left( 1+{x\over\rho}\right)\right) y' +
\left({\varepsilon -\lambda s\over x} +\zeta-\lambda{1+s\over\rho}\right)
y=0
\end{equation}
--with $\delta$, $\varepsilon$ and $\zeta$ as in \eqref{EIKOSITRIA}--
is reduced to a {Schr\"{o}dinger} equation
which must therefore be solvable, since Eq. \eqref{EIKOSITESSERA} is solvable.
More specifically, the expression $w(x)=c_1(x)/a(x)$ implies that 
the weight function in the relevant {Liouville} equation
(Eq. \eqref{PENTE}) will 
be equal to 
\begin{equation}\label{EIKOSIPENTE}  
\def\GONIES{\begin{array}{ll}
  \raisebox{.1cm}
 {${\scriptstyle\underline{\;\,x\to 0\,\;}}$} &\;
 \ln x\Rightarrow x=e^z \\ \noai\noa\noa \noalign{\smallskip}
\raisebox{.1cm}
{${\scriptstyle\underline{\;x\to \infty}}$} &\;
{1\over\sqrt{\rho \omega}}
\ln x\Rightarrow x=e^{\sqrt{\rho\omega} z}
\end{array}}
w(x)={1+(x/\rho)\over x^2(1+\omega x)}\Rightarrow
z(x)=\!\!\int\!\!\sqrt{w(x)}\; dx = \Bigg< \hskip-.15cm \GONIES
\end{equation}
where we have assumed $\rho$ and $\omega$ to be positive so that the 
change of variables $z=z(x)$ maps the region $0<x<\infty$ --between
the singular points $x=0$ and $x=\infty$ of the {initial equation}--
into the full region
$-\infty <z<\infty$ of the {\em final position variable} $z$.
As for the complete expression of the function  $z=z(x)$, this is a
simple elementary function which,
neverthelles, does not seem to be invertible 
in terms of elementary functions even though the existence
of an inverse is guaranteed by the fact that $z'(x)>0$
in the region of positive $x$.
But just as in the case of {Natanzon} potentials, this poses
no problem in solving the  {Schr\"{o}dinger}
equation and calculating its spectral properties, since 
only the asymptotic forms of $x(z)$ are involved
in these calculations and these are explicitly 
known (Eq. \eqref{EIKOSIPENTE}). We note also that the initial variable
$x$ is much more suitable for working out the solution; so we will 
use this variable henceforth in the understanding that at the end 
one has to make the substitution  $x\to x(z)$.

Based on the preceding discussion, we can now readily construct the potentials
derived from \eqref{EIKOSITESSERA} by simply applying formulas \eqref{ENA} 
through \eqref{ENTEKA}. Let us begin with the {\em Schwartzian term}
$V_s(x)=\{ z, x\}/w(x)$ which is given by
\begin{equation}\label{EIKOSIEXI}
V_s(x)={\rho\over 16}
{4\omega^2x^4+(4\omega +12\omega^2\rho)x^3+
(3+18\omega\rho +3\omega^2\rho^2) x^2+ (12\rho +4\omega\rho^2) x+4\rho^2
\over (x+\rho)^3 (1+\omega x)}
\end{equation}
while for the full potential $V(x)=\big( v(x)/w(x)\big) +V_s(x)$  we get

\begin{equation}\label{EIKOSIEPTA}
V(x)={Ax^4+Bx^3+Cx^2+Dx+E\over (x+\rho)^3 (1+\omega x)}
\end{equation}
whereby
\begin{equation}\label{EIKOSIOKTO}
\def\TOPA{\begin{array}{ll} 
A =& \displaystyle{\left({9\over 4}\rho\omega -3\right)\omega s^2 +
\left({3\over 2}\beta\rho -2\alpha -3\omega\rho +3\right)
\omega s+\left({\beta^2\rho\over 4\omega}+\rho\omega-\beta\rho\right)\omega}
\hfill\\ \noai\noa
B =& \displaystyle{\left({9\over 2}\omega^2\rho^2-{9\over 2}\omega\rho-3\right)
s^2+}\hfill\\ \noa
&\displaystyle{+\left(-{9\over 2}\omega^2\rho^2+3\omega\rho +3\omega\rho^2\beta
+{3\over 2}\rho\beta-{9\over 2}\alpha\omega\rho -2\alpha+3\right)s}
\hfill\\ \noai\noa
& \displaystyle{{1\over 2}\big(\rho^2\beta^2-3\omega\rho^2\beta 
+\rho\alpha\beta -3\alpha\omega\rho +3\omega\rho -\rho\beta\big)
+{1\over 4}\omega\rho (3\omega \rho +1)}\hfill\\ \noai\noa
C =& \displaystyle{{\rho\over 4}\bigg( 9(\omega^2\rho^2-3)s^2+
6(4-3\alpha -2\alpha\omega\rho +2\beta\rho +\omega\rho^2\beta -\omega^2\rho^2)s}
\hfill \\ \noa
& \hskip.7cm\displaystyle{
 +\rho^2\beta^2-2\omega\rho^2\beta +4\alpha\beta\rho 
- 2\beta\rho -10\alpha\omega\rho +\alpha^2-4\alpha} \hfill \\ \noa 
& \hskip1.3cm \displaystyle{
\left.
+ {3\over 4} \omega^2\rho^2+{9\over 2}\omega\rho +{15\over 4}\right)}
\hfill \\ \noa
D =& \displaystyle{{\rho^2\over 4} \big( 6(\omega\rho -3)s^2
+ (6\beta\rho -2\alpha\omega\rho -12\alpha +18)s} \hfill \\ \noa
&\hskip.7cm \displaystyle{
+ (2\alpha\beta\rho -4\alpha\omega\rho +2\alpha^2 -6\alpha +\omega\rho+3)
\big)}\hfill \\ \noa
E =& \displaystyle{\left(-{3\over 4}s^2-{1\over 2}
(\alpha -3)s +{1\over 4}(\alpha -1)^2\right) \rho^3 .}
\end{array}}
\TOPA
\end{equation}

Note that even though the potential is a 
{\em simple rational function} of the initial variable $x$, it 
nevertheless depends in a complicated way on the {\em five}
dimensionless parameters $\alpha$, $\beta$, $\omega$, $\rho$ and $s$
of the problem at hand. (There are {\em five} and not {\em six} parameters
as we had initially claimed, since the sixth is a {\em scaling parameter}
of $z$ which we implicitly set equal to unity --see, e.g., the first 
numerical coefficient in \eqref{EIKOSIDYO}-- so that, together with the
substitutions $\hbar =2m=1$, we can arrive at a {\em complete system of units} 
for the {Schr\"{o}dinger} equation.)
The form of $V(z)$ --which is but the form of $V(x)$ in the interval 
$0<x<\infty$ 
stretched to $-\infty <z<+\infty$-- will typically be as shown in 
Figure 1 with $V_0\,(=E/\rho^3)$ and $V_{\infty}\,(=A/\omega)$ given by
\begin{equation}\label{EIKOSIENNIA}
\def\ARTO{\begin{array}{ll} 
V_0 & \equiv V_{-\infty}=
{- \displaystyle{3\over 4}s^2-{1\over 2}
(\alpha -3)s+{1\over 4}(\alpha-1)^2} \\ \noa
V_{\infty} & =
{\displaystyle{\left({9\over 4}\rho\omega -3\right) s^2+
\left({3\over 2}\beta\rho -2\alpha -3\omega\rho +3\right) s
+{\beta^2\rho\over 4\omega}+\rho\omega -\beta\rho}}.  \end{array}}
\ARTO
\end{equation}
Even though the expression for the potential in terms of
its parameters is quite complex, the solutions
of the respective {Schr\"{o}dinger} equation
look much simpler! To a large extent, this simplification is due to
a different parametrization of the problem which occurs naturally
once we realize that Eq. \eqref{EIKOSITESSERA} has $x^s$ as a solution
provided $\lambda$ takes the value
\begin{equation}\label{TRIANTA}
\lambda_0=-\big( 3s^2+(2\alpha -3)s\big)
\end{equation}

\bigskip
\centerline{\psfig{file=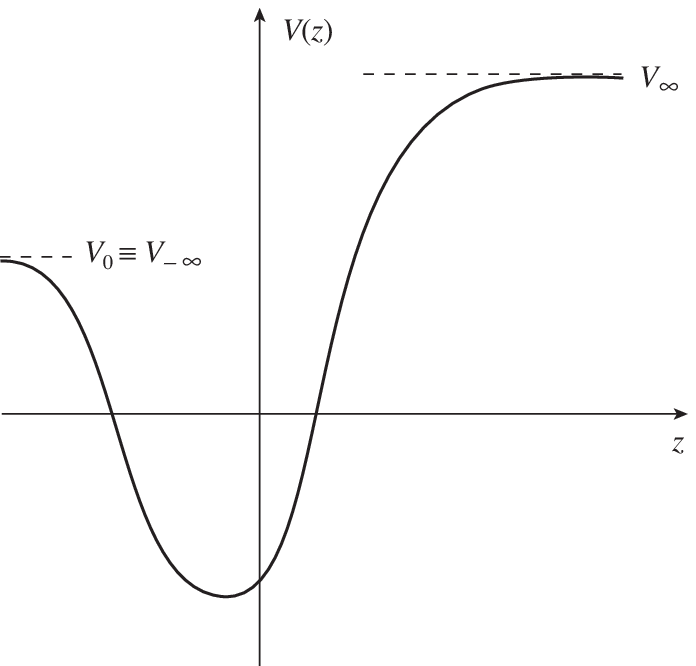,height=50mm}}

\smallskip
\centerline{\small {\bfseries Figure 1:}
A typical form of the potential $V(z)$}

\bigskip
\noindent
so that the corresponding eigenfunction $\psi_0 (x)$ is
\begin{equation}\label{TRIANTAENA}
\psi_0(x)={x^q(1+\omega x)^{r-q+(1/4)}\over (x+\rho)^{1/4}},
\end{equation}
where $q$ and $r$ are given by the expressions
\begin{equation}\label{TRIANTADYO}
q={3s+\alpha -1\over 2},\qquad r={3s+(\beta/\omega)-2\over 2}.
\end{equation}
But since for $x\to 0$ and $x\to\infty$ we have
$\psi_0(x)\to x^q$ and $\psi_0(x)\to x^r$ respectively,
it follows that \eqref{TRIANTAENA}
will satisfy the boundary conditions at the origin and at infinity 
--i.e., at $-\infty$ and $+\infty$ of the variable $z$--
only if $q>0$ and $r<0$. In this case, $\psi_0(x)$
represents the system's ground state since it has no nodes in this region. 
The crucial finding now is this: If we take the value \eqref{TRIANTA}
as the reference level for the potential energies and eigenvalues of the 
problem, using instead of $V$ and $\lambda$ the quantities
\begin{equation}\label{TRIANTATRIA}
U=V-\lambda_0,\qquad \epsilon =\lambda-\lambda_0
\end{equation}
then a remarkable simplification ensues. $U$ and $\epsilon$ do not 
depend separately on the parameters $s$, $\alpha$ and $\beta$, but only 
on their combinations $q$ and $r$. In other words, when we make this 
(parametrically dependent) change of reference level, the number of 
(dimensionless) parameters of the problem gets reduced from five to four
; $q$, $r$, $\rho$ and $\omega$. Note, for example, that subtracting 
 $\lambda_0$ from the asymptotic values \eqref{EIKOSIENNIA} gives 
\begin{equation}\label{TRIANTATESSERA}
U_0\equiv V_0-\lambda_0=q^2,\qquad 
U_{\infty}\equiv V_{\infty}-\lambda_0=\rho\omega r^2,
\end{equation}
which are much simpler expressions than those before, with the parameters 
now being $q,r,\rho$ and $\omega$. Let's call this way of parametrizing the
problem {\em invariant parametrization}; we shall work with this from now on.
The expression for the potential $U(x)$ now becomes as in \eqref{EIKOSIEPTA}, 
i.e., 
\begin{equation}\label{TREXI}
U(x)={{\cal A}x^4+{\cal B}x^3+{\cal C}x^2+{\cal D}x+{\cal E}
\over (x+\rho)^3(1+\omega x)}
\end{equation}
but with new coefficients 
${\cal A}$, ${\cal B}$, ${\cal C}$ etc.,
which are given by 
\begin{equation}\label{TREFT}
\def\FTATR{\begin{array}{rl}
{\cal A} = &\omega^2\rho r^2\hfill\\ \noa
{\cal B} = & 2\omega^2\rho^2r^2+2\omega\rho qr-\omega\rho q
+\omega^2\rho^2 r+\displaystyle{1\over 4}\omega\rho (1-\omega\rho)\hfill\\ \noa
{\cal C} = & \displaystyle{\rho 
\left( q^2+\omega^2\rho^2r^2+4\omega\rho qr-
(1+\omega\rho)q +\omega\rho (1+\omega\rho) r+{3\over 16}
(1-\omega\rho)^2\right)} \hfill\\ \noa
{\cal D} = & \displaystyle{\rho^2\left( 2q^2+2\omega\rho qr-q+\omega\rho r
+{1\over 4}(\omega\rho -1)\right)} \hfill\\ \noa
{\cal E} = & \rho^3q^2 . \end{array}}
\FTATR
\end{equation}.

\bigskip
\noindent
{\small\bfseries {IV}. RESULTS: THE FULL HYPERGEOMETRIC CASE}

\medskip

As explained in Appendix A,
if we use the above parametrization, the solution
$\psi(x)$ of the {Schr\"{o}dinger} equation satisfying the boundary 
condition $\psi(x\!=\!0)\!\equiv\! \psi(z\!=\!-\infty)\!=\!0$
in the discrete spectrum region
--or $\psi(z\to -\infty)\sim e^{ikz}$
in the case of continuous spectrum-- can be written as 
\begin{equation}\label{TRIANTAEPTA}
\psi(x) ={x^{\sqrt{q^2-\epsilon}} (1+\omega x)^{r-q+(1/4)}\over
(x+\rho)^{1/4}} \Big( x(x+\rho)F'+\big( ax+\rho (a+1)\big) F\Big)
\end{equation}
where $F\equiv$$_3F_2(a,b,c;d,e; -\omega x)$ and
$a,b,c,d,e$ are given by 
\begin{equation}\label{TRIANTAOKTO}
\def\TRIOK{\begin{array}{ll}
a = -q+\sqrt{q^2-\epsilon} & d = 2-q+\sqrt{q^2-\epsilon} =a+2\hfill\\ \noa\noai
b = \sqrt{q^2-\epsilon}+\sqrt{r^2-g\epsilon} -\sigma \qquad &
e = 1+2\sqrt{q^2-\epsilon} \hfill \\ \noa\noai
c = \sqrt{q^2-\epsilon} -\sqrt{r^2-g\epsilon} -\sigma\hfill\\ \noa\noai
\end{array}}
\TRIOK
\end{equation}
and
\begin{equation}\label{TRIANTAENNIA}
\sigma =q-r-1,\qquad g={1\over\rho\omega}.
\end{equation}
In fact it can be shown that expression \eqref{TRIANTAEPTA} can be
also written in the equivalent and more explicit form
$$\psi(x)={x^{\sqrt{q^2-\epsilon}}(1+\omega x)^{r-q+(1/4)}\over 
(x+\rho)^{1/4}}\times \hskip7cm $$
\begin{equation}\label{ANAXAL}
 \hskip.7cm \times \big(\rho (a+1)\, _3F_2 (a,b,c;a+1,e;-\omega x) +
ax\, _3F_2 (a+1,b,c; a+2, e; -\omega x)\big) 
\end{equation}
where no derivatives of $_3F_2$ enter. To the best of my knowledge
expressions like \eqref{ANAXAL} --with higher-order hypergeometric
functions present in the solution of a Schr\"{o}dinger equation--
appear for the first time in the literature.

To impose the boundary condition at $x\to +\infty$ (i.e., at $z\to +\infty$) 
we also need the asymptotic relation
\noindent
$$F(x)\mathop{\longrightarrow}\limits_{x\to\infty}
{\Gamma (d)\Gamma (e)\over\Gamma (b)\Gamma (c)}\,
{\Gamma (b-a)\Gamma (c-a)\over\Gamma (d-a)\Gamma (e-a)}(\omega x)^{-a}+
{\Gamma (d)\Gamma (e)\over\Gamma (a)\Gamma (c)}\,
{\Gamma (a-b)\Gamma (c-b)\over\Gamma (d-b)\Gamma (e-b)}(\omega x)^{-b}$$

\begin{equation}\label{SARANTA}
+{\Gamma (d)\Gamma (e)\over\Gamma (a)\Gamma (b)}\,
{\Gamma (a-c)\Gamma (b-c)\over\Gamma (d-c)\Gamma (e-c)} (\omega x)^{-c}
\end{equation}
in conjuction with the fact that the first-order operator 
${\cal M}$ acting on $F$\break --Eq. \eqref{TRIANTAEPTA}--
annihilates the term $x^{-a}$, for the same reason that 
$M$ annihilates also the solution $y_0$. Thus, we come to the following 
conclusions.

\bigskip
\noindent
{\small\bfseries {A}. Scattering states}

\smallskip

Denoting by $r_R(k,k')$ and $r_L(k,k')$ the reflection 
amplitudes from the right and left respectively, we have

\noindent
$$r_{R}(k,k')=
\displaystyle{r+i\sqrt{g}k'\over r-i\sqrt{g} k'}\,
\displaystyle{\Gamma (2i\sqrt{g}k') \over
\Gamma(-2i\sqrt{g}k')}
\times\hskip5.8cm$$
\begin{equation}\label{PENINTADYO}
\times\, {\Gamma(-\sigma-i(k+\sqrt{g}k'))\Gamma(\sigma+1-i(k+\sqrt{g}k'))
\over \Gamma(-\sigma-i(k-\sqrt{g}k'))\Gamma(\sigma+1-i(k-\sqrt{g}k'))}
\, \omega^{2i\sqrt{g} k'}
\end{equation}

$$r_{L} (k,k')= {-q+ik\over q+ik}\times
{\Gamma(1 +2ik)\over \Gamma(1-2ik)} \times\hskip5.9cm$$
\begin{equation}\label{PENINTAEXI}
\times\, {\Gamma (-\sigma -i(k+\sqrt{g}k'))
\Gamma (\sigma +1-i(k+\sqrt{g}k')) \over
\Gamma (-\sigma+i(k-\sqrt{g}k'))
\Gamma (\sigma +1+i(k-\sqrt{g}k'))} \omega^{-2ik}\hskip.5cm
\end{equation}

\smallskip
\noindent
where $k$ and $k'$ are the wavenumbers for  
$z\to -\infty$ and $z\to +\infty$ respectively.
In other words, we have $k=\sqrt{\epsilon -q^2}$, $k'=\sqrt{\epsilon -(r^2/g)}$.
For the corresponding reflection probability $P_r$ 
--which is independent of the direction of incidence-- we get
\begin{equation}\label{PENINTATRIA}
P_r(k,k')={\cosh^2\pi (k-\sqrt{g}k')-\cos^2\pi\sigma \over
\cosh^2\pi (k+\sqrt{g} k')-\cos^2\pi\sigma} .
\end{equation}

\noindent
{\small\bfseries {B}. Bound states}

\medskip
In this case, the requirement for vanishing $\psi (x)$ at infinity leads 
--due to \eqref{SARANTA}--
to the three conditions 
{i)} $a=\!-n$,
{ii)} $b=\!-n$ and
{iii)} $e-c=\!-n$. From these we obtain (see Appendix A for 
details) the respective regions,
{\em red}, {\em green} and {\em blue}, of the phase 
diagram in Figure 2. More specifically, the condition $a=-n$ 
is actually restricted 
to the value $n=0$ and has a respective (unique) eigenvalue 
$\epsilon =0$; while $b=-n$ and $e-c=-n$ lead to the equations (Ι)
and (ΙΙ) for the green and blue region respectively.

Note however that
in order for solutions of, say, Eq. (Ι) to exist, it is not sufficient
that $\sigma>0$, but we must also have $\sigma >f_{\min}$, where $f_{\min}$
is the minimum value of the function
$f(\epsilon)=\sqrt{q^2-\epsilon} +\sqrt{r^2-g\epsilon}$, in the left
hand side of (Ι), which is monotonically decreasing and will therefore
cross the horizontal line at height
$\sigma$ only if the above inequality holds.
This directly implies that the $g$-interval
within which bound states can exist is
\begin{equation}\label{Neosarantass}
{r^2-\sigma^2\over q^2}<g<
{r^2\over q^2-\sigma^2}
\end{equation}
where it is assumed that
the left or the right side of these
inequalities will be replaced by zero or infinity,
respectively, when the corresponding factor
$r^2-\sigma^2$ or $q^2-\sigma^2$
vanishes or becomes negative. It also follows from \eqref{Neosarantass}
that a suitable range of values for $g$ always exists, so the existence of
bound states in the green or blue region of the phase diagram
is always guaranteed.
Of course, the inequality \eqref{Neosarantass}
holds also for $\sigma\to\sigma -n$
--i.e., for the full right hand side of (Ι) $($or of (ΙΙ)$)$--
in which case the number of bound states will depend on the value of
$g$ and will generally be smaller than predicted by the
simple positivity of
the right hand sides of (Ι) or (ΙΙ).
Combining the above reasoning with
the fact that it will always be $q-r-1< |q|+|r|$, leads us to conclude
also that $\epsilon =0$ is not a solution of the eigenvalue equation (I),
although it can be
a solution of (ΙΙ) in the middle blue region; but ultimately even that is
rejected for reasons explained in Appendix A.
Hence the only vanishing eigenvalue
is the one located in the red region
and it represents the absolute ground state of the system.
In the special case whereby
the potential's asymptotic values $U_0$ and $U_{\infty}$ are equal
--i.e., when $q^2=r^2/g$-- then clearly
$f_{\min}=f(\epsilon)|_{\epsilon =q^2}=0$
and (Ι) is satisfied for all positive $g$,
a fact that also follows directly from the inequality
\eqref{Neosarantass}.
In that case, the eigenvalues are found easily in closed form and
their formula (in the green region, say) becomes



\GRAMMH
\medskip

\centerline{\psfig{file=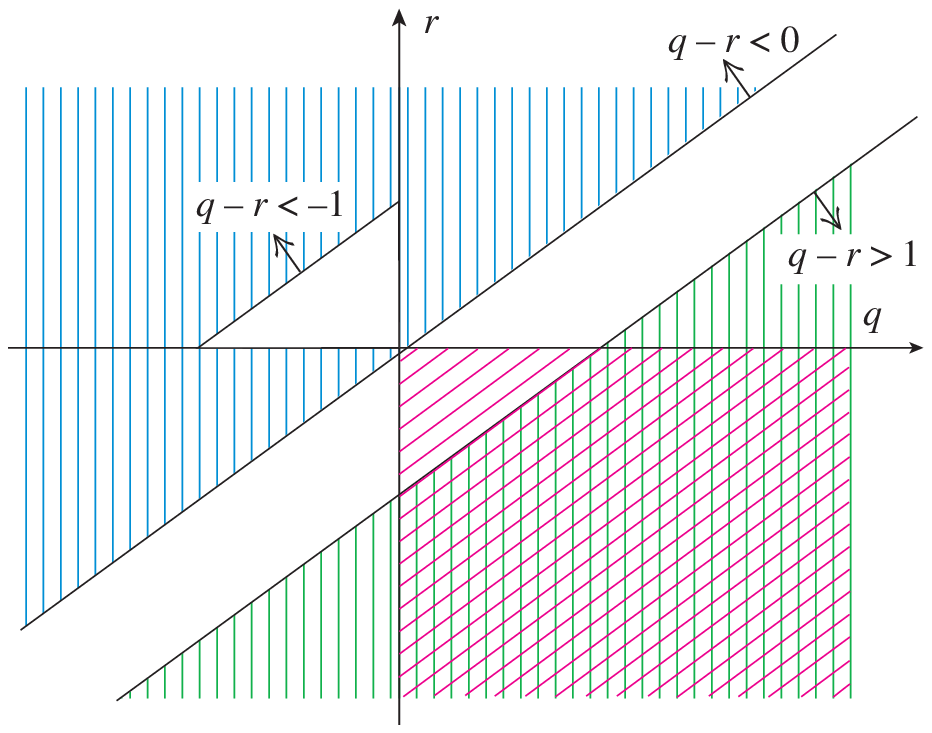,height=64mm}}  

\centerline{\small{\bfseries Figure 2:} Phase diagram for bound states
}
\centerline{\small in the full hypergeometric case.}

\medskip

{\small
\noindent
\hangindent=.3cm \hangafter=1
$\bullet\;${\bf Red region:} $(q>0, r<0)$. The eigenvalue 
$\epsilon =0$ exists and represents the absolute ground state of 
the system.

\noindent
\hangindent=.7cm \hangafter=1
$\bullet\;${\bf Green region:} $(q-r>1)$. Bound states exist and are 
determined by the condition

\medskip
{\hfill
$\sqrt{q^2-\epsilon}+\sqrt{r^2-g\epsilon}=\sigma -n= q-r-N\quad
(N=n+1)$
\hfill (I)}

\medskip
\noindent
\hangindent=.3cm \hangafter=1
$\bullet\;${\bf Blue region:} ($q-r<0$ for $qr>0$ or $q-r<-1$ for $qr<0$).
Bound states exist and are determined by the condition

\medskip
{\hfill$\sqrt{{q}^2-\epsilon}+\sqrt{{r}^2-g\epsilon}=
-\sigma -1-n = r-q-n$ \hfill(II)}

\medskip
\noindent
\hangindent=.3cm \hangafter=0
which can be also deduced from (I) by the substitution
$q\to -q$, $r\to -r$ and $N\to n$.
Note, however, that in 
the middle blue region $(qr<0)$ the value $n=0$ is rejected since 
then the value $\epsilon =0$, which is now a solution of (II),
does not correspond to a physically
acceptable solution (see Appendix A). 


\smallskip
\noindent
\hangindent=.7cm \hangafter=1
$\bullet\;${\bf White regions:} There are no bound states.

}

\vskip-.2cm

\GRAMMH

\bigskip
 \medskip
\noindent
\begin{equation}\label{SARANTAOKTO}
\epsilon_n=q^2-{(\sigma -n)^2\over(1+\sqrt{g})^2} =
q^2-{(q-r-N)^2\over (1+\sqrt{g})^2},\;\; N=n+1=1,2,\ldots\leqq q-r.
\end{equation}
This will also hold in the blue region upon substituting 
$q\to -{q}$, $r\to -{r}$;
while the further substitution $N\to n$ is also necessary in the 
outer blue regions.

Let us also mention that the wavefunction of the absolute ground state of the system 
-- we say {\em absolute} since there now exist  
various ground states across the different regions of the phase diagram
--is given by
\begin{equation}\label{Neoeikosiexi}
\psi_0(x)={x^q(1+\omega x)^{r-q+(1/4)}\over (x+\rho)^{1/4}} .
\end{equation}

Other features of the phase diagram worth noting are the following: 

a) In contrast to all known examples of solvable potentials, in this 
case the region of parametric space where bound states exist is 
{\em doubly connected}. It consists of the two main regions 
$q-r>1$ and $q-r<0$ with a gap in-between (white regions) where no 
bound states exist. 
b) Every point $(q,r)$, for example in the green-red region, has a 
{\em mirror image}  $(-q,-r)$ in the middle blue region with the
same bound states except the zero one ($\epsilon =0$) which exists in the green 
but not the blue region. This follows directly from the
eigenvalue equations (I) and (II) and the fact that in the middle blue region
the value  $n=0$ $(\Rightarrow \epsilon =0)$ is rejected. Thus we have 
pairs of potentials  --$U_+=U(q,r,x)|_{q-r>1}$ and $U_-=U(-q,-r,x)$-- 
with the same bound states except for $\epsilon =0$ which exists for the 
former but not the latter potential. Whether there is some
symmetry behind this ``pairing'' of potentials in parametric space 
is an interesting question that deserves further study.

\bigskip
\smallskip
\noindent
{\footnotesize\bfseries {V}. CONFLUENT CASE
OF THE FIRST KIND}
\smallskip

Let us now study the case where the parameter $\omega$ vanishes, 
so the solutions (see Appendix A) are hypergeometric functions of 
the confluent kind --i.e., $_2F_2$, $_1F_2$ etc. 
-- and their behavior at infinity will vary accordingly, including now
not only powers of $x$ but also exponentials. We will start
with the confluent case of the first kind, for which 
$\omega =0$ but $\beta\ne 0$. The expression of the potential in 
this case emerges readily from \eqref{TREXI} and  \eqref{TREFT} upon taking the 
limit $\omega \to 0$, which yields
\begin{equation}\label{enaA}
U(x)={{\cal A}x^4+{\cal B}x^3+{\cal C}x^2+
{\cal D}x+{\cal E}\over (x+\rho)^3}
\end{equation}
with
\begin{equation}\label{enaB}
\def\SSHI{\begin{array}{rl}
{\cal A}=& \displaystyle{1\over 4}\rho\beta^2\hfill\\ \noai
{\cal B}=& \displaystyle{1\over 2}\rho^2\beta^2+\rho q\beta\hfill\\ \noai
{\cal C}=& \displaystyle{\rho \left( q^2+{1\over 4}\rho^2\beta^2
+2\rho q\beta-q+{1\over 2}\rho\beta+{3\over 16}\right)}\hfill\\ \noai
{\cal D}=& \displaystyle{\rho^2\left( 2q^2+\rho q\beta -q+{1\over 2}
\rho\beta -{1\over 4}\right)}\hfill\\ \noai
{\cal E}=& \rho^3 q^2 \end{array} }
\SSHI
\end{equation}
and has a typical plot as in Figure 3. 

\bigskip
\centerline{\psfig{file=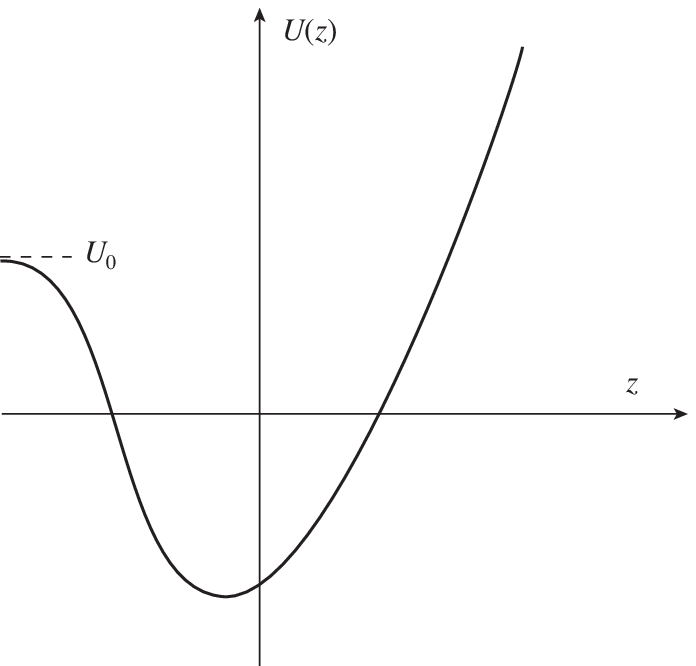,height=50mm}} 

\noindent
{\small {\bfseries Figure 3:}
Typical potential of the confluent family of the first kind. At $-\infty$ 
the potential tends to a {\em fixed value} and at $+\infty$ it rises as 
a harmonic oscillator.}

\medskip
\bigskip

\noindent
Let us note here that some of the results for the confluent case derive
directly from the previous ones by setting $\omega =0$, provided they 
do not represent a qualitative change in the problem,
in which case it is safer to redo the calculation.
Let us simply mention these results without elaborating much on them. 
We begin from the relation
$z=z(x)$, for which we have
\begin{equation}\label{PENINTAOKTO}
z'(x)={\sqrt{1+(x/\rho)}\over x}\;\Rightarrow\;
x(z) \mathop{\longrightarrow}\limits_{z\to -\infty} e^z, \qquad
x(z)\mathop{\longrightarrow}\limits_{z\to +\infty} {\rho\over 4}z^2.
\end{equation}
The general expression of the wavefunctions is now
\begin{equation}\label{PENINTAENNIA}
\psi(x)={x^{\sqrt{q^2-\epsilon}}e^{\beta x/2}\over (x+\rho)^{1/4}}
\Big( x(x+\rho)F'+\big( ax+\rho (a+1)\big) F\Big) ,
\end{equation}
where $F(x)\equiv {_2F_2}$ $(a,b;c,d;-\beta x)$ and
\begin{equation}\label{EXINTA}
\def\TNIX{\begin{array}{ll}
a=-q+\sqrt{q^2-\epsilon} & c=2-q+\sqrt{q^2-\epsilon}\;\; (=a+2) \\ \noa
b=-q+\sqrt{q^2-\epsilon} +p\epsilon +1\qquad & d=1+2\sqrt{q^2-\epsilon}
\end{array}}
\TNIX
\end{equation}
with $q$ as before --formula (30$_\mathrm{a}$)-- and
\begin{equation}\label{EXINTAENA}
p={1\over \rho\beta}
\end{equation}
where the new parameter $p$ takes now the position of $r$ in the phase
diagram that we will present shortly. 
The search for physically acceptable solutions depends crucially on 
the sign of $\beta$, or $p$, and is based on the asymptotic relation 
$$F(x)\mathop{\longrightarrow}\limits_{x\to\infty}
\displaystyle{\Gamma (b-a)\Gamma (c)\Gamma (d)\over
\Gamma (b)\Gamma (c-a)\Gamma (d-a)} (\beta x)^{-a} +
\displaystyle{\Gamma (a-b)\Gamma (c)\Gamma (d)\over
\Gamma (a)\Gamma(c-b)\Gamma(d-b)} (\beta x)^{-b}$$
\begin{equation}\label{EXINTADYO}
+ \displaystyle{\Gamma (c)\Gamma (d)\over\Gamma(a)\Gamma (b)}
(-\beta x)^{a+b-c-d} e^{-\beta x}.
\end{equation}
The results we obtain are as follows.

\bigskip
\noindent
{\small \bfseries {A.} Scattering states}

\medskip

Since the only scattering that makes sense now is from the left, 
we need only calculate the reflection amplitude 
$r_{L}(k)= r(k)$, which becomes
\begin{equation}\label{EXINTATRIA}
\def\XAAN{\begin{array}{ll}
r(k) = \displaystyle{-q+ik\over q+ik}\,
\displaystyle{\Gamma (1+2ik)\Gamma(1-q+p\epsilon-ik)\over
\Gamma (1-2ik)\Gamma (1-q+p\epsilon+ik)}(-\beta)^{-2ik}
\qquad & (\beta<0)\\ \noa
r(k) = \displaystyle{-q+ik\over q+ik}\,
\displaystyle{\Gamma (1+2ik)\Gamma(q-p\epsilon-ik)\over
\Gamma (1-2ik)\Gamma (q-p\epsilon+ik)}\beta^{-2ik}\qquad & (\beta>0)
\end{array}}
\XAAN
\end{equation}
with $|r(k)|=1$ of course, since the form of the potential does not allow
for the particle to pass through towards positive infinity. 


 \bigskip
\noindent
{\small \bfseries {B.} Bound states}

\medskip
A similar calculation as before yields now --apart from the vanishing 
eigenvalue in the region $q>0$, $p<0$, which is found separately (see
Appendix A)-- the two complementary conditions
\begin{equation}\label{Neopentess}
\sqrt{q^2-\epsilon} = -p\epsilon +q-N\qquad (p<0, N\geqq 1)
\end{equation}
and
\begin{equation}\label{Neopentepente}
\sqrt{q^2-\epsilon} ={p}\epsilon -{q}-n \qquad
(p>0, n\geqq 0)
\end{equation}
from which we can readily obtain the results summarized in the 
next phase diagram.

Clearly, the phase diagram has the type of {\em mirror symmetry}
we mentioned before and this raises similar questions as to its
origin and interpretation.

Concerning the absolute ground state of the system
--i.e. the one with $\epsilon =0$--
this exists only in the red region and its wavefunction is written as
\begin{equation}\label{PETRIA}
\psi_0(x)={x^q e^{\beta x/2}\over (x+\rho)^{1/4}}.
\end{equation}
Note also that the bound states do not depend on the full set of
dimensionless parameters $\rho$, $\beta$, $q$
of the potential $U(x)$ (in the invariant parametrization as before)
but only on their two combinations $p(=1/\rho\beta)$ and $q$.
This is contrary to what happens in the scattering data
--formulas \eqref{EXINTATRIA}-- where all three parameters enter.
Something similar to this occurs in the full hypergeometric case
we examined before. The presence of {\em isospectral orbits}
in parametric space is thus one more feature of the potentials presented
here which has to be explained.

\newpage

\noindent
\GRAMMH

\centerline{\psfig{file=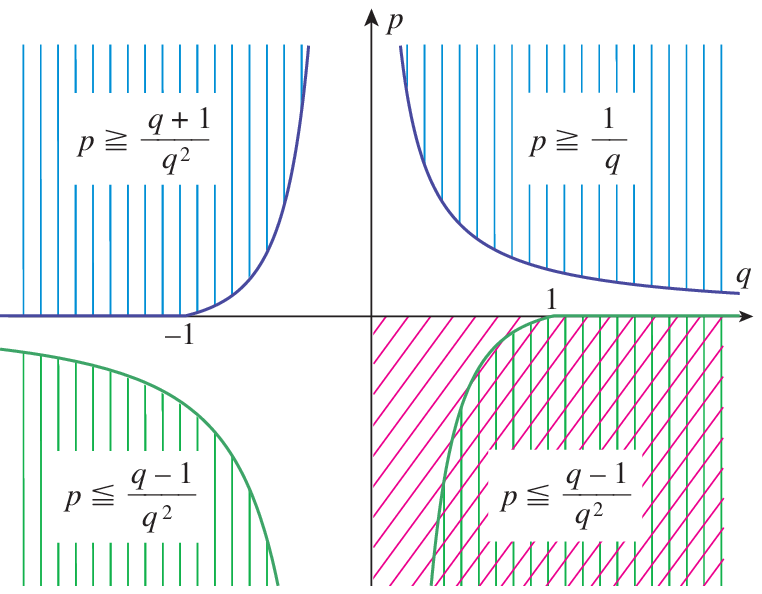,height=68mm}}

\centerline{\small {\bfseries Figure 4:}
Phase diagram for bound states}
\centerline{\small in the confluent case of the first kind.}

\medskip

{\small
\noindent
\hangindent=.3cm \hangafter=1
$\bullet\;${\bf Red region:} $(q>0, p<0)$.
The eigenvalue $\epsilon =0$ exists and represents the absolute
ground state of the system.

\noindent
\hangindent=.3cm \hangafter=1
$\bullet\;${\bf Green region:} $(p<0: p\leqq (q-1)/q^2)$.
Bound states exist, and they are given by the formula

\medskip
{\hfill$\displaystyle{
\epsilon_N={\left( pq-{1\over 2}\right)-pN+\sqrt{\left(pq-
{1\over 2}\right)^2+pN}\over p^2},\quad N=1,2,\ldots,\leqq (1-pq)q}$\hfill(I)}

\medskip

\noindent
\hangindent=.3cm \hangafter=1
$\bullet\;${\bf Blue region:} ($p>0: p\geqq 1/q$ for $pq>0$ and
$p\geqq (q+1)/q^2$ for $pq<0$). 
Bound states exist, and they are given by the formula

\smallskip
{\hfill$\displaystyle{
\epsilon_n={\left({pq}-{1\over 2}\right)+pn+
\sqrt{\left({p}{q}-{1\over 2}\right)^2-{p} n}
\over {p}^2},\quad n=0,1,\ldots\leqq
({p}{q}-1){q}}$ \hfill(II)}

\smallskip
\noindent
\hangindent=.3cm \hangafter=0
which can be also derived from (I) by the substitution
$p\to -p$, $q\to -q$ and $N\to n$.
But in the left blue region $(pq<0)$ the value $n=0$ 
is rejected for the same reasons as before.

\noindent
\hangindent=.3cm \hangafter=1
$\bullet\;${\bf White regions:} No bound states exist.

}
\vskip-.2cm

\GRAMMH

\bigskip

 \bigskip
\noindent
{\footnotesize\bfseries {VI.} CONFLUENT CASE OF THE SECOND KIND}
\medskip

The expression for the potential $U(x)$, in the invariant parametrization,
emerges directly from \eqref{enaA}, \eqref{enaB}
upon setting $\beta =0$, so there will be
\begin{equation}\label{EVDOMINTATRIA}
U(x)={\displaystyle{\left( q^2-q+{3\over 16}\right)\rho x^2 +
\left( 2q^2-q-{1\over 4}\right) \rho^2x+q^2\rho^3}\over (x+\rho)^3}
\end{equation}
with $x=x(z)$ as in \eqref{PENINTAOKTO}, which implies an exponentially
fast approach to the asymptotic value $U_0=q^2$ of the potential at 
$z\to -\infty$, but a very slow decrease to zero --of the kind  
$U\sim 1/z^2$-- at $z\to +\infty$. Depending on the range of values of 
$q$ the potential $U(z)$ can take one of the following forms: 

\bigskip
\smallskip
\centerline{
\psfig{file=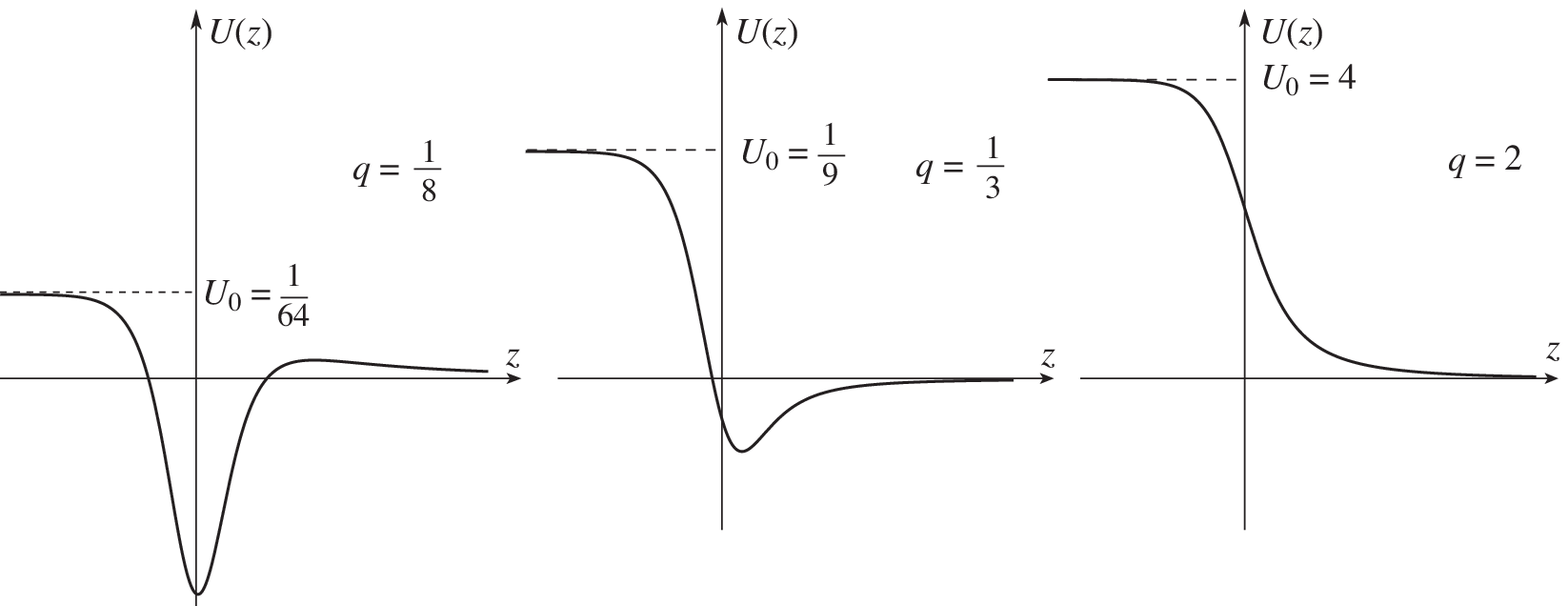,height=45mm}}

\smallskip
\centerline{\small {\bfseries Figure 5:}
Representative forms of the potential $U(z)$}
\centerline{\small in the confluent case of the second kind.}

\bigskip
The wave functions will now be written as
(Eq. \eqref{PENINTAENNIA} with $\beta =0$)
\begin{equation}\label{PSPSPS}
\psi (x)={x^{\sqrt{q^2-\epsilon}}\over (x+\rho)^{1/4}}
\big( x(x+\rho)F'+(ax+\rho (a+1))F\big)
\end{equation}
where $F(x)=\, _1F_2(a;b,c; -\epsilon x/\rho)$ and
\begin{equation}\label{PSBPSBPS}
a=-q+\sqrt{q^2-\epsilon},\qquad b=a+2,\qquad c=1+2\sqrt{q^2-\epsilon}
\end{equation}
while the asymptotic behavior of $F(\zeta)$ for large $|\zeta|$
--where $\zeta =-\epsilon x/\rho$ in our case--
will be given by
$$F(\zeta)\xrightarrow [|\zeta|\to\infty]{}
{\Gamma(b)\Gamma(c)\over\Gamma(b-a)\Gamma (c-a)}(-\zeta)^{-a} \hskip6.9cm$$
\begin{equation}\label{PSCPSCPSS}
+{\Gamma(b)\Gamma(c)\over 2\sqrt{\pi} \Gamma (a)}(-\zeta)^{\eta}
\Big( e^{i(\pi\eta +2\sqrt{-\zeta})}
+e^{-i(\pi\eta +2\sqrt{-\zeta})}\Big)\quad 
\left( \eta ={1\over 2}\left( a-b-c+{1\over 2}\right)\right)
\end{equation}
e.g., by a linear combination of the three asymptotic behaviors
at infinity, which now
have the form of two exponentials and one power
of $\zeta$ due to the fact that the maximum dimension
component of the operator ${\cal L}$ in \eqref{EIKOSITESSERA}
--with $\omega =\beta =0$-- is now of first order.

On the basis of the above the results we obtain are as follows.

For $\epsilon >U_0$ there are scattering solutions both from the 
left and from the right, and the corresponding reflection amplitudes
are given by the formulas
\begin{equation}\label{EVDOMINTAPENTE}
r_{L}(k)={-q+ik\over q+ik}\,
{\Gamma (1+2ik)\over \Gamma (1-2ik)}
\left({\epsilon\over\rho}\right)^{-2ik} e^{-2k\pi},
\quad r_{R} (k)=ie^{-2k\pi}
\end{equation}
where it is noteworthy that, apart from a constant phase (independent of $k$),
the reflection amplitude from the right is {\em real}.

As for bound states, there is only the eigenvalue $\epsilon =0$
with a corresponding eigenfunction
\begin{equation}\label{EVDOMINTAEXI}
\psi_0(x)={x^q\over (x+\rho)^{1/4}}
\end{equation}
which will satisfy the boundary conditions  
$\psi (0)=\psi(\infty)=0$ only if $0<q<1/4$.
Nevertheless, \eqref{EVDOMINTAEXI} is not square integrable
as one might have expected, due to the position of the corresponding
eigenvalue at the threshold of the continous spectrum.

It hardly needs mentioning, that all the above results 
--irrespective of how they were produced--
can be verified directly with a straight substitution in the 
 {Schr\"{o}dinger} equation. To do that, it is advisable 
to cast that equation in the equivalent $x$-form
\begin{equation}\label{EI}
{1\over w(x)}\psi''(x)-{w'(x)\over 2w^2(x)}\psi'(x)
+ \big(\epsilon -U(x)\big) \psi (x)=0
\end{equation}
where the derivatives and the expressions of the functions  
$w$ and $U$ are with respect to the initial variable $x$.

Finally, we note that in both main classes of potentials
(full hypergeometric case and confluent of the first kind) their 
graphs do not always assume the typical forms of Figures 
1 and 3 but they show interesting variations similar to those in 
Figure 5.  The question of which of these forms may have some 
special physical interest will not be discussed here.

\medskip
\noindent
{\footnotesize\bfseries {VII.} PENDING ISSUES AND GENERALIZATIONS}

\smallskip
Let us note first that our preceding discussion on potentials
that are solvable via hypergeometric functions of third order is
not exhaustive. The first reason for this is that the solution 
$y_0=x^s(x+\rho)$, which we have chosen to reduce the order
of the initial equation, is just a
{\em special case} --for $s' =1$-- of the general expression,
$y_0=x^s(x+\rho)^{s'}$, for the solutions of the first order equation
$My=0$. This special case is undoubtedly the most important, for reasons
that may have become obvious by now. But there are also other possibilities
--e.g., $s'=2$-- belonging to the {\em terminating series} category and 
which can thus be solutions of $Ly=0$, as required.
Given though that the coefficients of the relevant polynomial are now related to one
another (they are all functions of $\rho$), the number of independent 
parameters of the operator $L$, and hence of the potential $V$, will 
necessarily be reduced, or such a solution might not even exist. 

But there is another reason why our previous discussion is incomplete. 
We assumed that $\rho$ and $\omega$ are positive
--and hence, they fall outside the region $0<x<\infty$ between the 
singular points of the original equation--
which means that the potential $V$ is also finite in this region, and 
therefore also in $-\infty\!<\!z\!<\!+\infty$ 
into which $0<x<\infty$  is mapped 
via the transformation $z=z(x)$. But if we assume, for example, that 
$\omega <0$ --and substitute $\omega$ for $-\omega$-- then $x$ must lie
within the interval $0<x<\omega^{-1}$, which is mapped via the transformation
\eqref{EIKOSIPENTE} to the seminfinite $z$-interval $-\infty <z<0$; and so the 
corresponding potential $V(z)$  will also be a
half-interval potential
with repulsive core of the type $1/z^2$, for $z\to 0$. In the same fashion 
one needs to examine all the remaining choices of signs for $\rho$ and 
$\omega$ together with the corresponding choices of intervals between singular 
points with respect to $x$ or $z$. What is certain, is that even in the 
context of the third-order case, the set of solvable potentials is much wider
than what we presented here and deserves further study, especially if
such a study could provide answers to some interesting special questions. 

A third pending issue pertains to the comparison of the present family of
solvable potentials with that of {Natanzon}. What are the 
differences or similarities between the two families? Can one family be deduced 
from the other by applying the special {Darboux} transformation
related to the removal of the ground state? 
We will leave this discussion for a follow-up publication, once we 
have constructed a much more general framework of studying solvable potentials
and developed a general formalism that is immediately applicable to
{Natanzon} potentials as the simplest special case. 
What is certain is that  --apart from the special case $\rho =\omega^{-1}$ 
(for which the potentials are elementary functions)--
there is no overlap between the two families, as evidenced both from the 
different scattering data and the different functional forms of their solutions. 
(Hypergeometric functions of the type $_3F_2$, $_2F_2$, $_1F_2$
in our case, vs. $_2F_1$, $_1F_1$ in the case of {Natanzon}.)

Next comes the question whether our methodology can be 
extended to equations of order higher than third, and produce 
the corresponding families of solvable potentials. While this
will be the topic of an upcoming publication, we find it warranted
at this point to present, without proof, some of the pertinent
conclusions in order to underscore the existence of a general 
framework for the systematic investigation and enlisting of, potentially, 
all solvable potentials. Our most basic results are contained 
in the following three propositions: 

{\em{\bf\em Proposition 1}: Every formal eigenvalue equation of the 
general form 
\begin{equation}\label{ENENIDATESSERA}
(L+\lambda M)y=0
\end{equation}
where $L$ is any linear differential operator of order $n$
and $M$ a similar operator of order
\break 
$n-2$, can be reduced 
to an eigenvalue equation of second order 
--and hence, to a {Schr\"{o}\-din\-ger} equation--
provided that each solution of $My=0$ is also a solution of $Ly=0$.}

{\em{\bf\em Proposition 2}: The weight function $w(x)$
of the {Liouville} equation
{that results from the reduction of}
any given bidimensional equation $(L+\lambda M)y=0$
--subject to the constraints of Proposition 1--
can always be written in the form
\begin{equation}\label{EKEKA}
w(x)={1+(x/\rho)\over x^2(1+\omega x)}.
\end{equation}
Therefore, the transformation function  $z(x)$ that renders the
{Liouville} equation into {Schr\"{o}\-dinger} form
will also be a ``universal function'', i.e., the same function for all solvable
potentials. }

\smallskip
{\em{\bf\em Proposition 3:} The solvable potentials arising
from a bidimensional equation of an arbitrary order $(L+\lambda M) y=0$
--subject to the constraints of Proposition 1--
will always have the functional form
\begin{equation}\label{epEKA}
V(x)={Ax^4+Bx^3+Cx^2+Dx+E\over (x+\rho)^3(1+\omega x)}
\end{equation}
where the coefficients $A$, $B$, $C$ etc., are functions of the parameters
that remain free in the operators $L$ and $M$ once we demand that every 
solution of  $My=0$ be also a solution of  $Ly=0$}.

\smallskip
In subsequent publications, we will rely on these three propositions 
--and on some crucial extensions of them--
in order to extend this work and develop a systematic theory of exact 
solvability of the {Schr\"{o}dinger} equation.

 \bigskip

\noindent
{\footnotesize\bfseries {VIII}. DISCUSSION}
\smallskip

As was noted at the outset, the critical new idea of this work is this:
Given that the 
{Natanzon} search for exactly solvable potentials is {\em exhaustive},
we can only produce new ones if we enlarge the set of 
functions within which their solutions are sought for.
And if this extended
set were {\em the hypergeometric functions of all orders},
then utilizing this set for solving {Schr\"{o}dinger} equation
would be in principle possible only in conjunction 
to the idea of {\em order-reduction}. We are thus led naturally to the notion of 
{\em eigenvalue equations of higher order --of the hypergeometric type}
if we want them to be solvable-- that can be reduced to the 
{Schr\"{o}dinger} equation using the technique of 
{\em order-reduction}. And as we saw earlier, this idea is indeed feasible 
and has actually delivered as a tangible product a new 
set of solvable potentials, beyond that of {Natanzon}.
We also saw that this notion
--provided that propositions 1 through 3 are valid--
is extensible to a very elegant general formulation which encompasses, as 
special cases, both the one studied here 
--where $L$ was of third and $M$ of first order--
as well as the case of {Natanzon},
with $L$ being of second and $M$ of zeroth order,
i.e., a function of $x$.

It follows that the method is in principle extensible to equations 
of higher than third order, even though it remains to be seen whether 
the relevant calculations continue to be manageable. Many further questions 
arise; here is but a short list of these: 

a) Could there exist an {\em infinite hierarchy}
--an {\em infinite ``tower''}-- of solvable potentials,
with those of the {Natanzon} class occupying its ``ground level'', 
the ones we presented here the ``first floor'' and so on, 
for the potentials originating from equations of ever higher order? 
Or does the process terminate leading to an ultimate closed set
of solvable potentials? 

b) What is the origin of the peculiar topology of the phase diagram 
for bound states and its {\em mirror symmetry}? Are the potential pairs 
$U_+$ and $U_-$, connected by this symmetry, supersymmetric partners?
And what about the property of {\em shape invariance} that was shown 
not to match the initial expectations pertaining to its range of applicability?
In view of the preceding discussion, could it be that an old 
conjecture$^3$, namely that this invariance may require a wider 
class of potentials than that of {Natanzon},
ought to be investigated further?

c) What is the origin of the reparametrization that reduces the number 
of active parameters of the system, or of the isospectral orbits in 
parametric space? Do these properties hint at some sort of symmetry that 
allows also a purely algebraic approach to the problem?

At a more mathematical level, the mapping of eigenvalue equations
of higher order into the {Schr\"{o}dinger} eigenvalue equation
--a mapping that can be cast in an elegant compact form--
could potentially provide valuable new insights to an 
old equation which clearly continues to hold an element of surprise.

\bigskip
\bigskip
\centerline{\bfseries\small APPENDIX Α}
\centerline{\bfseries\small CALCULATION OF EIGENVALUES
AND EIGENFUNCTIONS}

\medskip
According to the equations \eqref{ENA} through \eqref{ENNIA} of section 
ΙΙ, we have
$$\psi (x)=(x+\rho)^{7/4} x^q(1+\omega x)^{r-q+(1/4)}
\left({y\over x^s(x+\rho)}\right)' \eqno \hbox{(A1)}$$
where $q$ and $r$ are defined as in  \eqref{TRIANTADYO} and
$y(x)$ is the solution of the bidimensional equation \eqref{EIKOSITESSERA}.
The latter can be written in terms of the hypergeometric function 
$_3F_2$ as
$$y(x)=x^{\mu}\, _3F_2(a,b,c;d,e; -\omega x) \eqno \hbox{(A2)}$$ 
where the factor $-\omega$, in the argument $\zeta=-\omega x$,
serves to transfer the
finite singular point $x_0=-\omega^{-1}$ 
of \eqref{EIKOSITESSERA} to the standard position $\zeta_0=1$
of the hypergeometric equation, while the factor 
$x^{\mu}$ represents the physically acceptable power behavior
of \eqref{EIKOSITESSERA} for
\break
$x\to 0$.
The possible values of $\mu$ (physically acceptable or not)
are found as solutions of the 
{\em unidimensional equation} (or {\em {Euler}-type equation})
$${\cal L}_1 x^{\mu} =0\eqno \hbox{(A3)}$$
where ${\cal L}_1$ is the lowest-dimension component of the 
bidimensional operator ${\cal L}=L+\lambda M$ of \eqref{EIKOSITESSERA}.
For the power behavior $x^{\nu}$ at infinity
the respective equation is
$${\cal L}_2x^{\nu}=0 \eqno \hbox{(A4)}$$
where ${\cal L}_2$ is the largest-dimension component 
of the operator ${\cal L}$. (Α3) and (Α4) are of course cubic equations for 
$\mu$ and $\nu$, which are however easily solved since we already know
(from the known solution $y_0=x^s(x+\rho)$)
that one behavior at the origin is $x^s$ and one behavior at 
infinity is $x^{s+1}$.
The exponents of the remaining two behaviors are easily found to be
$$\mu,\overline{\mu}={3-\alpha -s\over 2}\pm\sqrt{V_0-\lambda}, \qquad
\nu,\overline{\nu}={2-(\beta/\omega)-s\over 2} \mp
{1\over\sqrt{\rho\omega}} \sqrt{V_{\infty}-\lambda} \eqno \hbox{(A5)}$$
where the specific choice of signs for $\mu$ and $\nu$
(positive for $\mu$ and negative for $\nu$)
reflects the fact that
the physically desirable behaviors are then $x^{\mu}$ and $x^{\nu}$
for small and large $x$ respectively. Now we can readily calculate the 
hypergeometric parameters $a,b,c,d$ and $e$ in (Α2), by recalling that 
the parameters $a_i$ and $b_j$ in the general hypergeometric series 
$$_pF_q (a_1,a_2,\ldots, a_p; b_1,b_2,\ldots,b_q; z) =
\sum {(a_1)_n\ldots (a_p)_n\over (b_1)_n\ldots (b_q)_n}\,
{z^n\over n!} \eqno \hbox{(A6)}$$
have always the same {\em asymptotic meaning}:
$x^{-a_i}$ $(i=1,\ldots, p)$ are the power behaviors of the general 
solution of the corresponding hypergeometric equation at infinity and
$x^0$, $x^{1-b_j}$ $(j=1,\ldots,q)$ are the power behaviors at the origin.
The calculation of $a,b,c,d$ and $e$ in (Α2) reduces then to 
simple asymptotic comparison of the two sides for 
$x\to 0$ and $x\to\infty$. The result is
$$\begin{array}{c}a=\mu -s-1,\qquad b=\mu -\nu,\qquad c=\mu-\overline{\nu} 
\\ \noa
d=\mu -s+1,\qquad e=\mu-\overline{\mu}+1 \end{array}
\eqno \hbox{(A7)}$$
from which, upon using (Α5) and the definition of the new parameters 
$q$ and $r$ (eq. \eqref{TRIANTADYO}), we obtain 
the expressions \eqref{TRIANTAOKTO} and the general formula 
\eqref{TRIANTAEPTA} for the solution $\psi(x)$ that satisfies the 
boundary condition $\psi(0)=0$ in the region of bound states. 

The boundary condition at infinity ($\psi(\infty)=0$) is 
imposed via the asymptotic form of 
 $\psi(x)$ for large $x$ which is written as 
$$\psi(x)\to (a-b) Bx^{-\sqrt{r^2-g\epsilon}}+(a-c)Cx^{\sqrt{r^2-g\epsilon}}
\eqno \hbox{(A8)}$$
and is derived from the asymptotic relation \eqref{SARANTA} for $F(x)$
$$F(x)\equiv\, _3F_2\to Ax^{-a}+Bx^{-b}+Cx^{-c} \eqno \hbox{(A9)}$$
by taking into account what we noted earlier: 
Namely that the first-order operator
${\cal M}$ acting on $F$
(Eq. \eqref{TRIANTAEPTA}) annihilates the term $x^{-a}$.
It follows then from (Α8)
that the boundary condition at infinity 
will only be satisfied if $a-c=0$ or $C=0$. But the case 
$a-c=0$ is rejected at once due to the factor 
$\Gamma (a-c)$ in the expression for $C$, so we are left with 
$C=0$ which can be realized in each of the following ways 
$$\hbox{{i})}\;a=-n\qquad \hbox{{ii})}\;b=-n\qquad
\hbox{{iii})}\;e-c=-n \eqno \hbox{(A10)}$$
since the case $d-c=-n$ is also rejected due to the relation $d=a+2$
and the presense of the factor $\Gamma (a-c)$ in the numerator of $C$.
From $\Gamma (d)/\Gamma (a)=a(a+1)$ it is also clear that 
for case {(i)} only the values $a=0$ and $a=-1$
survive, the former reproducing the already known ground-state solution 
$\epsilon =0$ (i.e., $\lambda =\lambda_0$) for $q>0$, $r<0$,
while the latter gives  $\psi\equiv 0$ as expected.
In case {(ii)} the pertinent condition is written as 
$$\sqrt{q^2-\epsilon} + \sqrt{r^2-g\epsilon} =q-r-N \qquad
(N=1,2,\ldots) \eqno \hbox{(A11)}$$
and will be satisfied under the conditions laid out in 
section IV (Eq. \eqref{Neosarantass})
from where the exact number of bound states for the problem is deduced. 
Note that $\epsilon =0$ is no solution of (Α11)
for $n=0$, i.e., for $N=1$,
and thus the condition {(ii)} yields only positive eigenvalues.
In both the above cases, the hypergeometric function 
$F(x)$ has {\em polynomial form} because
both conditions $a=-n$ and $b=-n$ are {\em termination conditions}
for the respective hypergeometric series.

However, in case (iii) 
--contrary to what usually happens for bound states--
{\em the hypergeometric series does not
terminate} since in that case both behaviors 
$x^{-a}$ and $x^{-b}$ survive at infinity, even though only the latter
contributes to the asymptotic limit of $\psi (x)$.
The condition for calculating the eigenvalues is now written as 
$$\sqrt{{q}^2-\epsilon}+\sqrt{{r}^2-g\epsilon}= r-q-n
\qquad (n=0,1,2,\ldots,\leqq r-q) \eqno \hbox{(A12)}$$
but especially in the middle blue 
region of the relevant phase diagram (Fig. 2) the value $n=0$ has to 
be exempted since then (Α12) can have as its solution the vanishing 
eigenvalue $\epsilon =0$, which, however, does not lead to a physically 
acceptable solution $\psi(x)$ and must be thus rejected. The reason 
for this is nontrivial. For $\epsilon =0$ a flaw arises in the 
mechanism that ``ejects'' the undesirable asymptotic behavior 
for $x\to\infty$, since then the
hypergeometric parameters $a$, $b$, etc., no longer depend on $\rho$
(see Eq. \eqref{TRIANTAOKTO}),
so the corresponding solution 
$F(x)$ has no way of ``knowing'' the value of $\rho$
although the relevant {\em ejection operator} 
${\cal M}=x(x+\rho)\partial +ax+\rho(a+1)$
in \eqref{TRIANTAEPTA} continues to depend on it.

Using the same techniques as above --but with the asymptotic formula
\eqref{EXINTADYO} instead of \eqref{SARANTA}--
we can also calculate the bound states in the confluent
case, obtaining as a result 
--except for the zero eigenvalue--
the two conditions
$$\sqrt{q^2-\epsilon} = -p\epsilon +q-N,\qquad
\sqrt{q^2-\epsilon} =p\epsilon -q-n \eqno \hbox{(A13)}$$
which lead to formulas (Ι) and (ΙΙ) of Figure  4.

As for the scattering solutions
(in the full hypergeometric case for example),
we need the asymptotic form of $\psi (z)$
for $z\to\pm\infty$ which turns out to be 
$$\def\IGONI{\begin{array}{ll}
\raisebox{.1cm}
{${\scriptstyle\underline{\;z\to -\infty}}$} &\;\,
{ \rho^{3/4}(a+1)e^{ikz}}\\ \noai\noa\noa \noalign{\smallskip}
\raisebox{.1cm}
{${\scriptstyle\underline{\phantom{1}z\to \infty\phantom{1}}}$} &\;\,
{ \overline{B} e^{-ik'z}+\overline{C}e^{ik'z} }
\end{array}}
\raisebox{-.04cm}
{$\psi(z) =\Bigg< $}\hskip-.1cm \IGONI \eqno \hbox{(A14)}$$
$$\big(\overline{B}=\omega^{-\sigma-(3/4)} (a-b)B,\quad
\overline{C}=\omega^{-\sigma-(3/4)}(a-c)C\big)$$
and it is clear from (Α14) that the scattering solution from the right 
equals $\psi^*$ --that is, $\psi_R=\psi^*$--
and therefore
$$r_R=\left({\overline{B}\over\overline{C}}\right)^* \eqno \hbox{(A15)}$$ 
from where --after a few steps of algebra-- we readily obtain 
the results \eqref{PENINTADYO} and \eqref{PENINTAEXI}. 
Finally, the calculation of scattering from the left is done by
forming the combination $c_1\psi+c_2\psi^*$ containing only a 
traveling wave $\exp (ik'z)$ at  $+\infty$, and proceeding as before. 
The spectral data for the confluent cases of first and second kind
are calculated in an analogous manner. 

Some closing remarks about the connection of hypergeometric
functions and bidimensional equations are in order. The connection 
originates from the fact that bidimensional equations lead to  
two-term recursion relations,
from which the general coefficient of the
series-solution is computed 
in closed form and it turns out to be exactly the same
as the general coefficient of the hypergeometric series (Α6). 
Thus the hypergeometric functions emerge naturally as 
{\em solutions of the bidimensional equations of step unity}.
The equation they satisfy --take, for instance, the function
 $_3F_2$--
will naturally have the form \eqref{DEKATESSERA}, but with 
$\omega =-1$ (so that the nonzero singular point lies in the 
standard position $x_0=1$) and $\varepsilon =0$,
so that one exponent of the power behavior at the origin 
can take the typical value zero as in the general hypergeometric 
series (Α6) that solves the relevant equation. 
As for the exact form of the remaining numerical coefficients of 
 \eqref{DEKATESSERA}, as functions of the hypergeometric parameters
$a,b,c,d$ and $e$, this is never needed since to obtain the solutions 
it suffices to know the power behaviors of $y(x)$ at zero and 
infinity, together with the 
{\em standard asymptotic meaning} of the hypergeometric parameters.

As for the hypergeometric functions of the confluent type 
--i.e, with $p<q+1$-- these emerge naturally from those 
bidimensional equations whose highest-dimension component 
 $L_2$ has no longer the full order of the equation, but is of 
lower order by one, two, three, etc., exactly as the value of  
$p$ compared to $q+1$. In this case, only $p$ behaviors at 
infinity --where $p$ now stands also for the order of $L_2$--
will be powers of $x$; the rest will be exponentials. 
This alters the character of the point at infinity 
that now becomes {\em irregular singular}.
As for the terminology, we have chosen the terms 
{\em confluent equation} (or {\em function}) {\em of the first kind}
if $p=q$, {\em confluent of the second kind} if $p=q-1$, etc.
To complete the picture, we note that 
{\em not only the bidimensional equations of step one can be solved
via hypergeometric functions, but also the bidimensional equations of any
step} $\ell$. The reason for this is simple. It has to do with the 
(rather obvious) fact that the transformations 
$t=x^m$ (change of independent variable)
and $y=x^{\mu}Y$ (change of dependent variable)
{\em preserve the bidimensional character of an equation}
and merely cause a change in its step,  
from $\ell$ to $\ell/m$, and a shift in its starting powers 
by $\mu$, respectively.

It thus follows that we will always be able to write the solutions of 
any bidimensional equation in the form 
$$y(x)=x^{\mu}\, _p\Phi_q (a_1,\ldots,a_p; b_1,\ldots,b_q;kx^{\ell})
\eqno \hbox{(A16)}$$
where $_p\Phi_q$ is the {\em general solution} of the hypergeometric 
equation with parameters $a_i$ and $b_j$, $\ell$ is the step of the given 
equation and $k$ the numerical coefficient that maps all 
singular points outside the origin 
(all of which lie on a circle in the complex plane) to the standard 
position $\zeta =1$ where $\zeta =kx^{\ell}$.
As for the parameters $\mu$ and $a_i$, $b_j$ in (Α16) these are obtained 
readily from the  {\em asymptotic comparison} of its sides at zero and 
at infinity, leading directly to the relations
$$\nu_i=\mu -\ell a_i,\qquad \mu_j=\mu +\ell (1-b_j) \eqno \hbox{(A17)}$$
where $\mu_j$ and $\nu_i$ are the power behaviors of the given 
equation at zero and infinity respectively,
while of course  $\mu$ must be equal to one of the  
$\mu_j$ (e.g. $\mu=\mu_1$).

It is clear from the above discussion that the hypergeometric functions and 
bidimensional equations emerge as the {\em natural framework}
for the systematic study of the problem of exact solvability of the 
{Schr\"{o}dinger} equation or any other equation for that matter. 
Our future work on this problem will lie within this systematic framework.


\bigskip
\noindent
{\footnotesize\bfseries ACKNOWLEDGMENTS}
\smallskip

I am grateful to Nick Papanicolaou for helpful discussions 
and critical comments and Manolis Antonoyiannakis for helpful
discussions and encouragement.

\bigskip
\noindent
{\footnotesize\bfseries BIBLIOGRAPHY}
\smallskip

\noindent
\hangindent=.3cm \hangafter=1
{$^1\;$G.A. Natanzon, Vestnik Leningrand Univ. {\bf 10}, 22 (1971);
Teor. Mat. Fiz. {\bf 38}, 146 (1979)}

\noindent
\hangindent=.3cm \hangafter=1
{$^2\;$G. Darboux, C.R. Acad, Sci. (Paris) {\bf 94}, 1456 (1882)}

\noindent
\hangindent=.3cm \hangafter=1
{$^3\;$F. Cooper, N. Ginocchio and A. Khare, Phys. Rev. D {\bf 36},
2458 (1987)}

\noindent
\hangindent=.3cm \hangafter=1
$^4\;$F. Cooper, A. Khare, U. Sukhatme, Physics Reports {\bf 251},
{\bf 267} (1995)

\end{document}